\documentclass[conference,a4paper]{IEEEtran} 

\usepackage{amsmath}
\usepackage{amsfonts}
\usepackage{amssymb}
\usepackage{color}
\usepackage{cite}
\usepackage{graphicx}
\usepackage{epstopdf}

\newcommand*{\diff}{\mathop{}\mathopen{\mathrm{d}}}

\newcommand{\SNR}{\mathsf{SNR}}

\newcommand{\Exp}{\mathsf E}
\newcommand{\expect}[1]{{\Exp}\left[#1\right]}

\newcommand{\Var}{\mathsf{Var}}
\newcommand{\variance}[1]{{\Var}\left[#1\right]}

\newcommand{\h}{\mathsf h}
\newcommand{\ent}[1]{{\h}\left(#1\right)}

\newcommand{\I}{I}
\newcommand{\mi}[2]{{\I}\left(#1 \, ; #2 \right)}
\newcommand{\micnd}[3]{{\I}\left(\left. #1 ; #2 \,\right| #3\right)}

\newcommand{\C}{\mathsf C}
\newcommand{\capacity}[1]{{\C}\left(#1 \right)}

\DeclareMathOperator{\erf}{erf}
\DeclareMathOperator{\erfc}{erfc}

\title{Lower Bound on the Capacity of Continuous-Time Wiener Phase Noise Channels}
\author{
\authorblockN{Luca Barletta and Gerhard Kramer}
\authorblockA{Institute for Communications Engineering\\
Technische Universit\"{a}t M\"{u}nchen\\
D-80333 Munich, Germany\\
\{luca.barletta, gerhard.kramer\}@tum.de}}

\begin{document}

\maketitle

\begin{abstract}
 A continuous-time Wiener phase noise channel with an integrate-and-dump multi-sample receiver is studied.
 A lower bound to the capacity with an average input power constraint is derived, and a high signal-to-noise ratio (SNR) analysis is performed.
 The capacity pre-log depends on the oversampling factor, and amplitude and phase modulation do not equally contribute to capacity at high SNR.
\end{abstract}

\section{Introduction}
\label{Sec:introduction}

Instabilities of the oscillators used for up- and down-conversion of signals in communication systems give rise to the phenomenon known
as \emph{phase noise}.
The impairment on the system performance can be severe even for high-quality oscillators, if the continuous-time waveform
is processed by long filters at the receiver side.
This is the case, for example, when the symbol time is very long, as happens when using orthogonal frequency division multiplexing.

 
Typically, the phase noise generated by oscillators is a random process with memory, 
and this makes the analysis of the capacity challenging.
 The phase noise is usually modeled as a Wiener process, as it turns out to be accurate in describing the phase noise statistic
 of certain lasers used in fiber-optic communications~\cite{FoschiniVannucci}. As the sampled output of the filter matched to the transmit filter
 does not always represent a sufficient statistic~\cite{BarlettaISIT2014, BarlettaCROWNCOM2014}, oversampling does help in achieving higher rates over the continuous-time 
 channel~\cite{Martalo2013, GhozlanISIT2013, GhozlanGLOBECOM}.
 
 To simplify the analysis, some works assume a modified channel model where the filtered phase noise does not consider amplitude fading, 
 and thus derive numerical and analytical bounds~\cite{Barletta2013, BarlettaTIT2014, GhozlanISIT2014, BarlettaITW2015}.

The aim of this paper is to give a capacity lower bound without
any simplifying assumption on the statistic of filtered phase noise. Specifically,
we extend the existing results for amplitude modulation, partly published in~\cite{GhozlanISIT2013}, and present new
results for phase modulation.
%
%

\emph{Notation:} Capital letters denote random variables or random processes. The notation $X_m^n = (X_m,X_{m+1},\ldots, X_n)$ with $n \ge m$
is used for random vectors. With ${\cal N}(0, \sigma^2)$ we denote the probability distribution of a real Gaussian 
random variable with zero mean and variance $\sigma^2$. The symbol $\stackrel{{\cal D}}{=}$ means equality in
distribution.

Given a complex random variable $X$, we use the notation $|X|$ and $\angle X$ to denote the amplitude and the phase of $X$,
respectively. The binary operator $\oplus$ denotes summation modulo $[-\pi,\pi)$.

The operators $\expect{\cdot}$, $\ent{\cdot}$, and $\mi{\cdot}{\cdot}$ denote expectation, differential entropy, and mutual information,
respectively.

\section{System model}
\label{Sec:system}
The output of a continuous-time phase noise channel can be written as
\begin{equation}
 Y(t) = X(t) e^{j\Theta(t)} + W(t),\qquad 0\le t\le T \label{eq:model}
\end{equation}
where $j=\sqrt{-1}$, $X(\cdot)$ is the data bearing input waveform, and $W$ is a circularly symmetric complex white Gaussian noise.
The phase process is given by
\begin{equation}\label{eq:Theta}
 \Theta(t) = \Theta(0) + \gamma\sqrt{T} B(t/T),\qquad 0\le t\le T,
\end{equation}
where $B(\cdot)$ is a standard Wiener process, i.e., a process characterized by the following properties:
\begin{itemize}
 \item $B(0)=0$,
 \item for any $1\ge t> s\ge 0$, \mbox{$B(t)-B(s)\sim {\cal N}(0,t-s)$} is independent of the sigma algebra generated by \mbox{$\{B(u): u\le s\}$},
 \item $B(\cdot)$ has continuous sample paths.
\end{itemize}
One can think of the Wiener phase process as an accumulation of white noise:
\begin{equation}\label{eq:ThetaWiener}
 \Theta(t) = \Theta(0) + \gamma \int_0^t B'(\tau)\diff \tau,\qquad 0\le t\le T,
\end{equation}
where $B'(\cdot)$ is a standard white Gaussian noise process.

\subsection{Signals and Signal Space}
\label{subsec:signals}
Suppose $X(\cdot)$ is in the set ${\cal L}^2[0,T]$ of finite-energy signals in the interval $[0,T]$.
Let $\{\phi_{m}(t)\}_{m=1}^\infty$ be an orthonormal basis of ${\cal L}^2[0,T]$. We may write
\begin{align}
X(t)= \sum_{m=1}^\infty X_m \: \phi_m(t),
\quad W(t) = \sum_{m=1}^\infty W_m \: \phi_m(t) \label{eq:W}
\end{align}
where
\begin{align}
X_m &= \int_{0}^{T} X(t) \: \phi_m(t)^\star \diff t,
\end{align}
$x^\star$ is the complex conjugate of $x$, and the $\{W_m\}_{m=1}^\infty$ are independent and identically distributed (iid), 
complex-valued, circularly symmetric, 
Gaussian random variables with zero mean and unit variance.

The projection of the received signal onto the $n-$th basis function is
\begin{align}
 Y_n &= \int_{0}^{T} Y(t) \: \phi_n(t)^\star \diff t\\
 &= \sum_{m=1}^\infty X_m \int_{0}^{T} \phi_m(t) \: \phi_n(t)^\star \: e^{j\Theta(t)} \diff t + W_n \\
 &= \sum_{m=1}^\infty X_m \: \Phi_{mn} + W_n. \label{eq:mimo}
\end{align}
The set of equations given by \eqref{eq:mimo} for $n=1,2,\ldots$ can be interpreted as the output of an infinite-dimensional multiple-input multiple-output channel, whose
fading channel matrix is $\Phi=[\Phi_{mn}]$.

\subsection{Receivers with Finite Time Resolution}
\label{Sec:receiver}
Consider a receiver whose time resolution is limited to $\Delta$ seconds, in the sense that every projection must
include at least a $\Delta$-second interval. More precisely, we set $ML\Delta=T$, where $M$ is the number of independent 
symbols transmitted in $[0,T]$ and $L$ is the oversampling factor, i.e., the number of samples per symbol.
The integrate-and-dump receiver with resolution time $\Delta$ uses the basis functions
\begin{align}
   \phi_m(t) = \left\{ \begin{array}{ll}
   1/\sqrt{\Delta}, & t \in [(m-1)\Delta,m\Delta) \\
   0, & \text{elsewhere}.
   \end{array} \right. \label{eq:rect}
\end{align}
for $m=1,\ldots,ML$.
With the choice~\eqref{eq:rect}, the fading channel matrix $\Phi$ is diagonal and the channel's output for \mbox{$n=1,\ldots,ML$} is
\begin{align}
 Y_n &= X_n \: \frac{1}{\Delta} \int_{(n-1)\Delta}^{n \Delta} e^{j\Theta(t)} \diff t+ W_n \nonumber\\
 &=X_n \: e^{j \Theta((n-1)\Delta)}\frac{1}{\Delta} \int_{(n-1)\Delta}^{n \Delta} e^{j(\Theta(t)-\Theta((n-1)\Delta))} \diff t+ W_n \nonumber\\
 &\stackrel{{\cal D}}{=} X_n \: e^{j \Theta_n}\frac{1}{\Delta} \int_{0}^{\Delta} e^{j\gamma \sqrt{\Delta} B_n(t/\Delta)} \diff t+ W_n \label{eq:Ync}\\
 &\stackrel{(a)}{=} X_n \: e^{j \Theta_n} \int_{0}^{1} e^{j\gamma \sqrt{\Delta} B_n(t)} \diff t+ W_n \nonumber\\
 &=X_n \: e^{j \Theta_n} F_n+ W_n,\label{eq:Yne}
\end{align}
where we have used the notation $\Theta_n=\Theta((n-1)\Delta)$ and $F_n = \int_{0}^{1} e^{j\gamma \sqrt{\Delta} B_n(t)} \diff t$.
In~\eqref{eq:Ync} we have used~\eqref{eq:Theta}, the property $B(t/T)-B((n-1)\Delta/T)\stackrel{{\cal D}}{=}B(t/T-(n-1)\Delta/T)$,
the substitution 
\begin{equation}
 \left\{\begin{array}{l}
	    t\leftarrow t-(n-1)/\Delta \\
	    B_n(t/T) \leftarrow B(t/T - (n-1)\Delta/T),
        \end{array}\right.
\end{equation}
and the property $\sqrt{T}B_n(t/T)\stackrel{{\cal D}}{=} \sqrt{\Delta}B_n(t/\Delta)$. Finally, in step~$(a)$ we have used the
substitution $t \leftarrow t/\Delta$.

Since the oversampling factor is $L$, we have $X_{kL+1}=X_{kL+2}=\ldots=X_{kL+L}$ for $k=0,\ldots,M-1$, and we can write the model~\eqref{eq:Yne}
as
\begin{equation}\label{eq:contmodel}
 Y_n = X_{\lceil n/L \rceil L} \: e^{j \Theta_n} F_n+ W_n
\end{equation}
for $n=1,\ldots, ML$.

The vectors $X_1^{ML}$, $F_1^{ML}$, and $W_1^{ML}$
are independent of each other. 
The variables $\{X_{kL}\}_{k=1}^M$ are chosen as iid with zero mean and variance
$\expect{|X_{n}|^2}$, and the average power constraint is 
\begin{align}
 \expect{\frac{1}{T}\int_0^T |X(t)|^2 \diff t} &= \frac{1}{ML\Delta}  \sum_{n=1}^{ML} \expect{|X_n|^2 } \nonumber\\
 &=\frac{\expect{|X_n|^2 }}{\Delta}\le {\cal P}. \label{eq:power}
\end{align}
Since we set the power spectral density of $W$ to 1, the power ${\cal P}$ is also the SNR, i.e., 
$\SNR = {\cal P}$.

Using~\eqref{eq:ThetaWiener}, the variables $\Theta_1^{ML}$ follow a discrete-time Wiener process:
\begin{equation}\label{eq:ThetaWiener1}
 \Theta_n = \Theta_{n-1} +  N_{n-1}, \qquad n=1,\ldots, ML,
\end{equation}
where the $N_n$'s are iid Gaussian variables with zero mean and variance $\gamma^2 \Delta$.
The fading variables $F_n$'s are complex-valued and iid, and $F_n$ is independent of $\Theta_1^{n}$.
In other words, $F_n$ is correlated only to $N_n$, and is independent of the vector $(N_1^{n-1},N_{n+1}^{ML})$.

Note that for any finite $\Delta$, or equivalently for any finite oversampling factor $L$, the vector $Y_1^{ML}$ does not
represent a sufficient statistic for the detection of $X$ given $Y$ in the model~\eqref{eq:model}.


\section{Lower bound on capacity}\label{Sec:upper}
We compute a lower bound to the capacity of the continuous-time Wiener phase noise channel~\eqref{eq:contmodel}-\eqref{eq:ThetaWiener1}.
For notational convenience, we use the following indexing for $i=1,\ldots, L$ and $k=1,\ldots, M$:
\begin{equation}\label{eq:model1}
 Y_{(k-1)L+i} = X_k \: e^{j \Theta_{(k-1)L+i}} F_{(k-1)L+i} + W_{(k-1)L+i},
\end{equation}
and we group the output samples associated with $X_k$ in the vector ${\bf Y}_k=Y_{(k-1)L+1}^{(k-1)L+L}$.

The capacity is defined as
\begin{equation}\label{eq:capacity}
 \capacity{\SNR} = \lim_{M\rightarrow \infty}\frac{1}{M}\sup \mi{X_1^M}{{\bf Y}_1^M}
\end{equation}
where the supremum is taken among the distributions of $X_1^M$ such that the average power constraint~\eqref{eq:power} is satisfied.

The mutual information rate can be lower-bounded as follows:
\begin{align}
 &\frac{1}{M}\mi{X_1^M}{{\bf Y}_1^M} = \frac{1}{M}\sum_{k=1}^M\micnd{X_k}{{\bf Y}_1^M}{X_1^{k-1}} \nonumber\\
 & \stackrel{(a)}{=}\frac{1}{M}\sum_{k=1}^M\micnd{|X_k|}{{\bf Y}_1^M}{X_1^{k-1}}+\micnd{\angle X_k}{{\bf Y}_1^M}{X_1^{k-1},|X_k|} \nonumber\\
 & \stackrel{(b)}{\ge}\frac{1}{M}\sum_{k=1}^M \mi{|X_k|^2}{||{\bf Y}_k||^2}+\micnd{\angle X_k}{{\bf Y}_{k-1}^k}{X_{k-1},|X_k|} \nonumber\\
 & = \underbrace{\mi{|X_1|^2}{||{\bf Y}_1||^2}}_{I_{||}} + \underbrace{\micnd{\angle X_1}{{\bf Y}_{0}^1}{X_{0},|X_1|}}_{I_{\angle}} \label{eq:first}
\end{align}
where step~$(a)$ follows by polar decomposition of $X_k$, step~$(b)$ holds by a data processing inequality, by reversibility of the map $x \mapsto x^2$ for non-negative reals, 
and because $X_1^{k-1}$ is independent of $(X_k, {\bf Y}_k)$. Finally, the last equality follows by stationarity of the processes.

\subsection{Amplitude Modulation}
By choosing a specific input distribution that satisfies the average power constraint we always get a lower bound on 
the mutual information,
so we choose the input distribution as
\begin{equation}
 \label{eq:input_distr}
 p_{|X_k|^2}(x) = \left\{\begin{array}{lr}
                     \frac{1}{\lambda} \exp\left(-\frac{x-\Delta^{-t} }{\lambda}\right) & x\ge \Delta^{-t} \\
                     0 & \text{elsewhere}
                    \end{array}\right.
\end{equation}
where $\lambda = \SNR \Delta-\Delta^{-t}>0$ with $t>0$. Note that with this choice the average power constraint is satisfied with equality, i.e., $\expect{|X_k|^2}=\SNR\Delta$.

Similar to the method used in~\cite{GhozlanISIT2013}, we give here a lower bound to the first term on the right hand side (RHS) of~\eqref{eq:first} in the form
\begin{equation}
 I_{||}\ge \expect{-\ln q_{V}(V)} - \expect{-\ln q_{V |\: |X_1|^2}(V |\: |X_1|^2)}
\end{equation}
where $V=||{\bf Y}_1||^2$ and
\begin{equation}\label{eq:aux_output}
 q_V(v) = \int_0^\infty p_{|X_1|^2}(x) q_{V |\: |X_1|^2}(v | x) \diff x.
\end{equation}
Specifically, we choose the \emph{auxiliary channel} distribution as
\begin{equation}\label{eq:auxiliary}
 q_{V |\: |X_1|^2}(v | x) = \frac{1}{\sqrt{\pi \nu x}} \exp\left(-\frac{(v-L(1+x\expect{G}))^2}{\nu x}\right)
\end{equation}
where $G=||{\bf F}_1||^2/L$ and $\nu>0$, for which we have\footnote{Details are provided in the extended version of the paper.}
\begin{align}\label{eq:V_given_X}
 &\expect{-\ln q_{V |\: |X_1|^2}(V |\: |X_1|^2)} = \frac{1}{2}\ln\left(\pi\nu\right)+\frac{1}{2} \expect{\ln(|X_1|^2)}\nonumber\\
 &\quad + \frac{L}{\nu}\left(\frac{\expect{|X_1|^2}}{\Delta} \variance{G}+  2\expect{G}+\expect{\frac{1}{|X_1|^2}}\right)\nonumber\\
 &\le \frac{1}{2}\ln\left(\pi\nu\lambda\right)+\frac{\Delta^{-t} }{2\lambda} + \frac{L}{\nu}\left(\SNR\cdot \variance{G}+  2+\Delta^t\right)
\end{align}
where the inequality is due to $\expect{G}\le 1$, $\expect{|X_1|^2}\le\SNR\Delta$, the bound $\expect{|X_k|^{-2}}\le \Delta^t$ which follows from the support of $|X_k|^2$, and
\begin{align}
 \expect{\ln|X_1|^2} &= \int_{\Delta^{-t}}^\infty \frac{1}{\lambda} \exp\left(-\frac{x-\Delta^{-t} }{\lambda}\right) \ln(x) \diff x \nonumber\\
 &= \ln\lambda + \int_{\Delta^{-t}/\lambda}^\infty \exp\left(-\left(u-\frac{\Delta^{-t} }{\lambda}\right)\right) \ln(u) \diff u \nonumber\\
 &\le \ln\lambda +\frac{\Delta^{-t} }{\lambda}. \label{eq:explogX}
\end{align}
 
By substituting~\eqref{eq:auxiliary} and~\eqref{eq:input_distr} into~\eqref{eq:aux_output}, and 
by following similar steps to those of~\cite{GhozlanISIT2013}, we get
\begin{equation}\label{eq:V}
 \expect{-\ln q_{V}(V)} \ge -\frac{\Delta^{-t} }{\lambda}+\frac{1}{2}\ln(L^2\mu^2\lambda^2+\lambda\nu).
\end{equation}
By putting together~\eqref{eq:V_given_X} and~\eqref{eq:V} we obtain
\begin{align}
 I_{||}&\ge -\frac{3\Delta^{-t} }{2\lambda}+\frac{1}{2}\ln(L^2\mu^2\lambda^2+\lambda\nu) - \frac{1}{2}\ln\left(\pi\nu\lambda\right)\nonumber\\
 &\quad - \frac{L}{\nu}\left(\SNR\cdot \variance{G}+  2+\Delta^t\right). \label{eq:amplitude_mi}
\end{align}

In the limit of large time resolution we have 
\begin{equation}\label{eq:averageG}
\lim_{\Delta\rightarrow 0} \frac{\variance{G}}{\Delta^3} = \frac{\gamma^2}{45}.
\end{equation}
Now we let the time resolution grow as a power of the SNR, i.e., $\Delta^{-1}=\lceil \SNR^\alpha \rceil$, and the parameter $\nu=\rho\Delta^{-\beta}$, with $\rho>0$. By using~\eqref{eq:averageG} into~\eqref{eq:amplitude_mi}, in order to find a tight bound in the interval $1/3\le\alpha\le 1$ we need to satisfy the conditions\footnote{Details are provided in the extended version of the paper.} $\alpha < 1/(t+1)$ and $\beta \ge 1$. The tightest bound is obtained with $\beta=1$ and $\rho=4$:
\begin{align}\label{eq:limit_amplitude1}
 \lim\limits_{\SNR\rightarrow\infty}\left\{ I_{||} - \frac{1}{2}\ln(\SNR) \right\} \ge -\frac{1}{2}\ln(4\pi e).
\end{align}
For $0<\alpha<1/3$ we need to satisfy the conditions $\alpha < 1/(t+1)$ and $\alpha  \ge 1/(\beta+2)$, and the tightest bound is obtained by choosing $\beta = \alpha^{-1}-2$ and $\rho = 2\gamma^2/45$:
\begin{align}\label{eq:limit_amplitude2}
\lim_{\SNR\rightarrow\infty}\left\{I_{||}-\frac{3\alpha}{2}\ln\left(\SNR\right)\right\}
&\ge  - \frac{1}{2}\ln\left(\frac{2\pi\gamma^2 e}{45}\right).
\end{align}

\subsection{Phase Modulation}
The second term in the RHS of~\eqref{eq:first} can be lower-bounded as follows
\begin{align}
&\micnd{\angle X_1}{{\bf Y}_{0}^1}{X_{0},|X_1|} \stackrel{(a)}{\ge} \micnd{\angle X_1}{\Phi}{X_{0},|X_1|}\nonumber\\
&\ge \expect{-\ln q_{\Phi|X_0,|X_1|}(\Phi|X_0,|X_1|)} -\expect{-\ln q_{\Phi|X_0^1}(\Phi|X_0^1)} \label{eq:phase}
\end{align}
where step~$(a)$ is due to a data processing inequality with 
\begin{align}\label{eq:processing_phase}
\Phi&=\angle(Y_1 (Y_0 e^{-j\angle X_0} )^\star) \nonumber\\
&=\angle X_1 \oplus \angle(|X_1|F_1+W_1) \oplus \angle(|X_0|F_0^\star e^{j N_0}+W_0^\star),
\end{align}
and the last inequality follows by choosing the auxiliary channel
\begin{equation}\label{eq:auxiliary_phase}
q_{\Phi|X_0^1}(\phi|x_0^1) = \frac{\exp(\zeta \cos(\phi-\angle x_1))}{2\pi I_0(\zeta)}
\end{equation}
where $I_0(\cdot)$ is the zero-th order modified Bessel function of the first kind, and $\zeta$ is a positive real number. Since we assume an uniform input  phase distribution, 
the output distribution is also uniform:
\begin{equation}\label{eq:uniform}
q_{\Phi|X_0,|X_1|}(\phi|x_0,|x_1|) = \int_0^{2\pi} q_{\Phi|X_0^1}(\phi|x_0^1) \frac{1}{2\pi} \diff \angle x_1 = \frac{1}{2\pi}.
\end{equation}
Using~\eqref{eq:auxiliary_phase}, the second term in the RHS of~\eqref{eq:phase} can be upper-bounded as follows for any $\Delta \le \bar\Delta<\infty$:
\begin{align}
\expect{-\ln q_{\Phi|X_0^1}(\Phi|X_0^1)} &= \ln(2\pi I_0(\zeta)) - \zeta\expect{ \cos(\Phi-\angle X_1)}\nonumber\\
&\le \ln(\pi\sqrt{\pi})+\frac{1}{2}\ln\left(\frac{1}{\zeta}\right)+\zeta \rho\nonumber\\
&= \frac{1}{2}\ln\left(2\pi^3 e\rho\right) \label{eq:phase_mod}
\end{align}
where the inequality is due to $I_0(\zeta)\le \sqrt{\pi}/2 \cdot e^\zeta/\sqrt{\zeta}$ derived in~\cite[Lemma 2]{GhozlanISIT2010}, and from the result of Appendix~\ref{App:cos} with
\begin{equation}
 \rho = 1-\expect{F_0 e^{-j N_0}}\expect{F_1}+2e^{-3\gamma^2\Delta/8}\expect{|X_1|^{-2}}K_{\bar\Delta}
\end{equation}
where $K_{\bar\Delta}>1$ is a finite number\footnote{For example, choosing $\gamma^2\Delta=0.01$ gives $K_{\bar\Delta}=8.1353$. See the extended version of the paper for a detailed derivation.}.
The last step in~\eqref{eq:phase_mod} is obtained by choosing $\zeta = (2\rho)^{-1}$.

In the limit of large time resolution we have 
\begin{equation}\label{eq:averageRho}
\lim_{\Delta\rightarrow 0} \left\{\frac{\rho}{\Delta}- 2 K_{\bar\Delta} \Delta^{t-1}\right\} \le \frac{2}{3}\gamma^2 
\end{equation}
where the inequality follows from the bound $\expect{|X_1|^{-2}}\le \Delta^t$. Choosing $t=1$ and putting together~\eqref{eq:phase} and~\eqref{eq:uniform}-\eqref{eq:averageRho}
we get
\begin{align}
 \lim_{\Delta\rightarrow 0} \bigg\{I_{\angle}+\frac{1}{2}\ln(\Delta)\bigg\} \ge \frac{1}{2}\ln\left(\frac{3}{\pi e (\gamma^2+3 K_{\bar\Delta})}\right),
\end{align}
and letting the time resolution grow as a power of the SNR, i.e., $\Delta^{-1}=\lceil \SNR^\alpha \rceil$, for $0<\alpha\le 1/2$ we have
\begin{align}\label{eq:limit_phase}
 \lim_{\SNR\rightarrow \infty} \left\{I_{\angle}-\frac{\alpha}{2}\ln(\SNR)\right\}\ge \frac{1}{2}\ln\left(\frac{3}{\pi e (\gamma^2+3 K_{\bar\Delta})}\right).
\end{align}

\section{Discussion}\label{Sec:discussion}

\begin{figure}
 \begin{center}
 \includegraphics[width=\columnwidth]{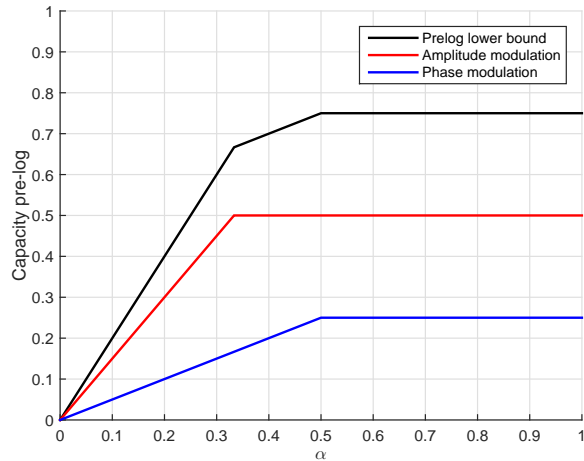}
\end{center}
\caption{Capacity pre-log lower bounds as a function of $\alpha$ at high SNR. The oversampling factor $L$ is $L=\lceil \SNR^\alpha \rceil$.}
\label{fig}
\end{figure}

As a byproduct of~\eqref{eq:limit_amplitude1}, \eqref{eq:limit_amplitude2}, and~\eqref{eq:limit_phase}, a lower bound to the capacity pre-log is
\begin{equation}\label{eq:prelog}
 \lim_{\SNR\rightarrow\infty} \frac{\capacity{\SNR}}{\ln(\SNR)} \ge 
 \left\{
 \begin{array}{ll}
  2\alpha & 0<\alpha\le 1/3 \\
  (1+\alpha)/2 & 1/3\le\alpha\le 1/2 \\
  3/4 & 1/2 \le \alpha < 1. 
 \end{array}\right.
\end{equation}
Figure~\ref{fig} shows the lower bounds on the capacity pre-log versus the parameter $\alpha$, as reported by~\eqref{eq:prelog}.
The contributions of amplitude and phase modulation are also shown separately: Amplitude modulation reaches full degrees of 
freedom by sampling more than $\sqrt[3]{\SNR}$ samples per symbol, while phase modulation achieves at least half of the available degrees
of freedom by using a time resolution that scales as $1/\sqrt{\SNR}$. 

The input distribution that achieves the capacity lower bound is uniform in phase and the square amplitude is distributed as a shifted exponential~\eqref{eq:input_distr}.
The statistic used for detecting $|X_k|$ is $||{\bf Y}_k||$, and the one used for detecting $\angle X_k$ is $\angle \left( Y_{(k-1)L+1} \left(Y_{(k-1)L}e^{-j\angle X_{k-1}}\right)^\star \right)$.

%
%
%
%

 \section{Conclusions}\label{Sec:conclusion}
 We have derived a lower bound to the capacity of continuous-time Wiener phase noise channels with an average transmit power constraint.
 As a byproduct, we have obtained a lower bound to the capacity pre-log at high SNR that depends on the growth rate
 of the oversampling factor used at the receiver. If the oversampling factor grows proportionally to $\SNR^\alpha$,
 then a capacity pre-log as high as that reported in~\eqref{eq:prelog} can be achieved.
%

\appendices
\section{A lower bound to $\expect{ \cos(\Phi-\angle X_1)}$}\label{App:cos}
The expectation can be simplified as follows:
\begin{align}
&\expect{ \cos(\Phi-\angle X_1)} \nonumber\\
&\stackrel{(a)}{=} \expect{ \cos( \angle(|X_1|F_1+W_1)-\angle(|X_0|F_0 e^{-j N_0}+W_0))}\nonumber\\
&\stackrel{(b)}{=} \expect{\cos(\angle(|X_1|F_1+W_1))}\expect{ \cos(\angle(|X_0|F_0 e^{-j N_0}+W_0))}\nonumber\\
&\ +\expect{\sin(\angle(|X_1|F_1+W_1))}\expect{ \sin(\angle(|X_0|F_0 e^{-j N_0}+W_0))}\nonumber\\
&=\expect{\cos(\angle(|X_1|F_1+W_1))}\expect{ \cos(\angle(|X_0|F_0 e^{-j N_0}+W_0))} \label{eq:app_expectation}
\end{align}
where step~$(a)$ is due to~\eqref{eq:processing_phase}, step~$(b)$ to the addition formula for cosine and independence of random variables, and the last step follows because $\expect{\sin(\angle(|X_1|F_1+W_1))}=0$ as $F_1$ and $W_1$ have symmetric pdfs with respect to the real axis. 

The first expectation on the RHS of~\eqref{eq:app_expectation} can be written as
\begin{align}
\expect{\Re\{e^{j\angle(|X_1| F_1+W_1)}\}}&=  \expect{\Re\{e ^{j\angle F_1}e^{j\angle(|X_1 F_1|+W_1)}\}}\nonumber\\
&= \expect{\cos(\angle F_1)\cos(\angle(|X_1 F_1| +W_1))} \label{app:first_expectation}
\end{align}
where the first step is due to the circular symmetry of $W_1$, and the second step because of the symmetric pdfs of $F_1$ and $W_1$.
A lower bound to~\eqref{app:first_expectation} is given by
\begin{align}
&\expect{\Re\{e^{j\angle(|X_1| F_1+W_1)}\}}\stackrel{(a)}{\ge} \expect{\Re\{F_1\}\cos(\angle(|X_1 F_1| +W_1))} \nonumber\\
&\stackrel{(b)}{\ge}\expect{ \Re\{F_1\}1(\Re\{F_1\}\ge 0)\left(1-\frac{1}{|X_1 F_1|^2}\right)} \nonumber\\
&\quad+\expect{\Re\{F_1\}1(\Re\{F_1\}< 0) }\nonumber\\
&\stackrel{(c)}{\ge} \expect{\Re\{F_1\} - 1(\Re\{F_1\}\ge 0)\frac{1}{|X_1 F_1|^2} }\nonumber\\
&\stackrel{(d)}{\ge} \expect{\Re\{F_1\} - \frac{1}{|X_1 F_1|^2} } \nonumber\\
&\ge \frac{2}{\gamma^2\Delta}\left(1-e^{-\gamma^2\Delta/2}\right) - \expect{\frac{1}{|X_1|^2}}  K_{\bar\Delta} \label{app:cos2}
\end{align}
where step~$(a)$ holds because $|F_1|\le 1$, $(b)$ follows by $\cos(x)\le 1$ and by\footnote{The proof is provided in the extended version of the paper.}
\begin{equation}
 \expect{\cos(\angle(\rho +W_1))} \ge 1-\frac{1}{\rho^2}, \qquad \rho>0,
\end{equation}
step~$(c)$ because $\Re\{F_1\}\le 1$, step~$(d)$
is obtained by subtracting $\expect{1(\Re\{F_1\}<0)\: |X_1 F_1|^{-2}}$, and the final inequality uses $\expect{|F_1|^{-2}}\le K_{\bar\Delta}$ for a finite suitable $\bar\Delta$ 
\footnote{The proof is provided in the extended version of the paper.}.

Following an analogous derivation used for finding~\eqref{app:cos2}, for the second factor on the RHS of~\eqref{eq:app_expectation} we have
\begin{align}
 &\expect{\Re\{e^{j\angle(|X_0| F_0 e^{-j N_0}+W_0)}\}}\ge \expect{ \Re\{F_0 e^{-j N_0}\} - \frac{1}{|X_0 F_0|^2} }   \nonumber\\
 &\ge \sqrt{\frac{2\pi}{\gamma^2\Delta}}\erf\left(\sqrt{\frac{\gamma^2\Delta}{8}}\right) e^{-3\gamma^2\Delta/8} - \expect{\frac{1}{|X_0|^2}}  K_{\bar\Delta} \label{app:cos3}
\end{align}
where $\erf(\cdot)$ is the error function, and the closed form for $\expect{F_0 e^{-j N_0}}$ is provided in Appendix~\ref{App:expect}. Using~\eqref{app:cos2} and~\eqref{app:cos3} into~\eqref{eq:app_expectation}, with $\Re\{\expect{F_1}\}\le 1$, $\Re\{\expect{F_0 e^{-j N_0}}\}\le e^{-3\gamma^2\Delta/8}$, 
$\expect{|X_0|^{-2}}\ge 0$, and $\expect{|F_0|^{-2}}\ge 0$, the final result is
\begin{align}
 \expect{ \cos(\Phi-\angle X_1)} &\ge \expect{F_0 e^{-j N_0}}\expect{F_1}\nonumber\\
 &\quad-2e^{-3\gamma^2\Delta/8}\expect{\frac{1}{|X_1|^2}}  K_{\bar\Delta}.
\end{align}

\section{Evaluation of $\expect{F_0 e^{-j N_0}}$}\label{App:expect}
Knowing that $N_0=\sigma\int_0^1 B(\tau)\diff \tau$ with $\sigma=\gamma\sqrt{\Delta}$, we compute
\begin{align}
 \variance{\sigma B(t)-N_0} &= \sigma^2\variance{B(t)}+\variance{N_0}-2\sigma\expect{B(t)N_0} \nonumber\\
 &=\sigma^2 (t + 1) -2\sigma^2 \int_0^1 \expect{B(t)B(\tau)} \diff \tau \nonumber\\
 &=\sigma^2(t^2-t+1)\label{app:variance}
\end{align}
where the last step follows from the property of Wiener processes $\expect{B(t)B(\tau)}=\min\{t,\tau\}$. Thus we have
\begin{align}
 \expect{F_0 e^{-j N_0}}&= \int_0^1 \expect{e^{j(\sigma B(t)-N_0)}} \diff t \nonumber\\
 &\stackrel{(a)}{=}\int_0^1 e^{-\variance{\sigma B(t)-N_0}/2} \diff t \nonumber\\
 &=\sqrt{\frac{2\pi}{\sigma^2}} e^{-\frac{3}{8}\sigma^2} \erf\left(\sqrt{\frac{\sigma^2}{8}}\right)
\end{align}
where in step~$(a)$ we used the characteristic function of a Gaussian random variable, and in the last step we used~\eqref{app:variance}.

\section*{Acknowledgment}
L. Barletta and G. Kramer were supported by
an Alexander von Humboldt Professorship endowed by the
German Federal Ministry of Education and Research.

\bibliographystyle{IEEEtran}
\bibliography{refSuffStat}

\newpage

\section{Auxiliary channel}\label{App:E}
Choose the auxiliary channel distribution
\begin{equation}\label{eq:app_auxiliary}
 q_{V |\: |X_1|^2}(v | x) = \frac{1}{\sqrt{\pi \nu x}} \exp\left(-\frac{(v-L(1+x\mu))^2}{\nu x}\right)
\end{equation}
where
\begin{align}
 V =||{\bf Y}_1||^2= |X_1|^2 LG + Z_0 + 2|X_1|Z_1
\end{align}
and
\begin{equation}
 G = \frac{||{\bf F}_1||^2}{L}, \qquad Z_0 = ||{\bf W}_1||^2,\qquad Z_1 = \sum_{i=1}^L\Re\{F_i W_i^\star\}.
\end{equation}
Squaring the output statistic gives
\begin{align}
 V^2 &= |X_1|^4 L^2 G^2 +4 |X_1|^3 LG Z_1  + 2|X_1|^2(2 Z_1^2 +  LG  Z_0) \nonumber\\
 &\quad+  4|X_1| Z_0 Z_1+ Z_0^2,
\end{align}
thus
\begin{align}
 &(V-L(1+|X_1|^2\mu))^2 \nonumber\\
 &= V^2 - 2VL(1+|X_1|^2\mu) + L^2(1+|X_1|^2\mu)^2 \nonumber\\
 &= |X_1|^4 L^2 G^2 +4 |X_1|^3 LG Z_1  + 2|X_1|^2(2 Z_1^2 +  LG  Z_0) \nonumber\\
 &\quad+  4|X_1| Z_0 Z_1+ Z_0^2 + L^2(1+|X_1|^2\mu)^2\nonumber\\
 &\quad -2(|X_1|^2 LG + Z_0 + 2|X_1|Z_1)L(1+|X_1|^2\mu)  \nonumber\\
 &= |X_1|^4 L^2 ( G -\mu)^2  + 2|X_1|^2(2 Z_1^2 +  L(Z_0 -L)( G- \mu))  \nonumber\\
 &\quad +4 |X_1|^3 L Z_1  (G  - \mu) +  4|X_1| Z_1 (Z_0  -L)+ (Z_0-L)^2.
\end{align}
Taking expectations gives
\begin{align}
 &\expect{\frac{(V-L(1+|X_1|^2\mu))^2}{\nu |X_1|^2}} \nonumber\\
 &= \frac{L^2}{\nu}\expect{|X_1|^2} \expect{( G -\mu)^2}+ \frac{4L}{\nu}\expect{|X_1|}  \expect{Z_1  (G  - \mu) }\nonumber\\
 &\quad+ \frac{2}{\nu}(2 \expect{Z_1^2} +  L\expect{(Z_0 -L)( G- \mu)})\nonumber\\
 &\quad +  \frac{4}{\nu}\expect{\frac{1}{|X_1|}} \expect{Z_1 (Z_0  -L)}+\frac{1}{\nu}\expect{\frac{1}{|X_1|^2}} \expect{(Z_0-L)^2}\nonumber\\
 &= \frac{L^2}{\nu}\expect{|X_1|^2} \expect{( G -\mu)^2}+  \frac{2L}{\nu}\expect{G}+\frac{L}{\nu}\expect{\frac{1}{|X_1|^2}}
\end{align}
which is minimized by choosing $\mu=\expect{G}$.

The conditional entropy is
\begin{align}
&\expect{-\ln q_{V |\: |X_1|^2}(V |\: |X_1|^2 )} \nonumber\\
&= \frac{1}{2}\ln(\pi\nu) +\frac{1}{2} \expect{\ln(|X_1|^2)} + \expect{\frac{(V-L(1+|X_1|^2\mu))^2}{\nu |X_1|^2}}\nonumber\\
&= \frac{1}{2}\ln\left(\pi\nu\right)+\frac{1}{2} \expect{\ln(|X_1|^2)}\nonumber\\
&\quad + \frac{L}{\nu}\left(\frac{\expect{|X_1|^2}}{\Delta} \variance{G}+  2\expect{G}+\expect{\frac{1}{|X_1|^2}}\right)\nonumber\\
&\le \frac{1}{2}\ln\left(\pi\nu\lambda\right)+\frac{\Delta^{-t} }{2\lambda} \nonumber\\
&\quad+ \frac{L}{\nu}\left(\SNR\cdot \variance{G}+  2\expect{G}+\expect{\frac{1}{|X_1|^2}}\right)
\end{align}
where in the last inequality we have used the bound~\eqref{eq:explogX}.

The output distribution is
\begin{align}
&q_V(v) = \int_0^\infty p_{|X_1|^2}(x) q_{V |\: |X_1|^2}(v | x) \diff x \nonumber\\
&= \int_{\Delta^{-t}}^\infty \frac{1}{\lambda} \exp\left(-\frac{x-\Delta^{-t} }{\lambda}\right)  \frac{1}{\sqrt{\pi \nu x}}\nonumber\\
&\qquad \exp\left(-\frac{(v-L(1+x\mu))^2}{\nu x}\right) \diff x \nonumber\\
&\le \exp\left(\frac{\Delta^{-t} }{\lambda}\right)\int_{0}^\infty \frac{1}{\lambda} \exp\left(-\frac{x}{\lambda}\right)  \frac{1}{\sqrt{\pi \nu x}}  \nonumber\\
&\qquad \exp\left(-\frac{(v-L(1+x\mu))^2}{\nu x}\right) \diff x \nonumber\\
&= \exp\left(\frac{\Delta^{-t} }{\lambda}\right)\int_{0}^\infty \frac{1}{L\mu\lambda} \exp\left(-\frac{y}{L\mu\lambda}\right)  \frac{1}{\sqrt{\pi \nu y/(L\mu)}}\nonumber\\
&\qquad \exp\left(-\frac{(v-L-y))^2}{\nu y/(L\mu)}\right) \diff y \nonumber\\
&= \exp\left(\frac{\Delta^{-t} }{\lambda}\right) \frac{1}{\sqrt{L\mu\lambda(L\mu\lambda+\nu/(L\mu))}}\nonumber\\
&\quad\exp\left(\frac{2L\mu}{\nu}\left[v-L-|v-L|\sqrt{1+\frac{\nu}{L^2\mu^2\lambda}}\right]\right),
\end{align}
and the entropy of the output of the auxiliary channel is
\begin{align}
&\expect{-\ln q_{V }(V  )} =-\frac{\Delta^{-t} }{\lambda}+\frac{1}{2}\ln(L\mu\lambda(L\mu\lambda+\nu/(L\mu)))\nonumber\\
&\quad-\frac{2L\mu}{\nu}\left[\expect{V-L}-\expect{|V-L|}\sqrt{1+\frac{\nu}{L^2\mu^2\lambda}}\right]\nonumber\\
&\stackrel{(a)}{\ge}-\frac{\Delta^{-t} }{\lambda}+\frac{1}{2}\ln(L\mu\lambda(L\mu\lambda+\nu/(L\mu)))\nonumber\\
&\quad+\frac{2L\mu}{\nu}\expect{V-L}\left[\sqrt{1+\frac{\nu}{L^2\mu^2\lambda}}-1\right]\nonumber\\
&\ge-\frac{\Delta^{-t} }{\lambda}+\frac{1}{2}\ln(L^2\mu^2\lambda^2+\lambda\nu).
\end{align}
where step $(a)$ holds because $\expect{|\cdot|}\ge \expect{\cdot}$, and the last inequality holds because of
\begin{align}
\expect{V-L}&=\expect{|X_1|^2 LG + Z_0 + 2|X_1|Z_1-L}\nonumber\\
&=\expect{|X_1|^2 || {\bf F}_1 ||^2 }\ge 0.
\end{align}
The lower bound to the mutual information rate for the amplitude modulation is
\begin{align}
I_{||}&\ge \expect{-\ln q_{V }(V  )} - \expect{-\ln q_{V |\: |X_1|^2}(V |\: |X_1|^2 )}\nonumber\\
&\ge -\frac{3\Delta^{-t} }{2\lambda}+\frac{1}{2}\ln(L^2\mu^2\lambda^2+\lambda\nu) - \frac{1}{2}\ln\left(\pi\nu\lambda\right)\nonumber\\
&\quad - \frac{L}{\nu}\left(\SNR\cdot \variance{G}+  2\expect{G}+\expect{\frac{1}{|X_1|^2}}\right).
\end{align}
Using $\lambda = \SNR\Delta-\Delta^{-t}$ and choosing $\nu = \rho\Delta^{-\beta}$, and $\Delta = L^{-1} = \SNR^{-\alpha}$ gives
\begin{align}
I_{||}
&\ge -\frac{3}{2}\frac{1 }{\SNR^{1-\alpha(t+1)}-1}+\frac{1}{2}\ln\left(\SNR^{2\alpha}(\expect{G})^2\right.\nonumber\\
&\left.(\SNR^{1-\alpha}-\SNR^{\alpha t})^2+\rho(\SNR^{1-\alpha(1-\beta)}-\SNR^{\alpha (t+\beta)})\right) \nonumber\\
&\quad- \frac{1}{2}\ln\left(\pi\rho\SNR^{1+\alpha(\beta-1)}\left(1-\frac{1 }{\SNR^{1-\alpha(t+1)}}\right)\right)\nonumber\\
&\quad - \rho^{-1}\SNR^{\alpha(1-\beta)}\left(\SNR\cdot \variance{G}+  2\expect{G}+\SNR^{-\alpha t}\right).
\end{align}
For large SNR we have
\begin{equation}\label{eq:app_averageG}
\lim_{\Delta\rightarrow 0} \expect{G} = 1,\qquad
\lim_{\Delta\rightarrow 0} \frac{\variance{G}}{\Delta^3} = \frac{\gamma^2}{45}
\end{equation}
that gives
\begin{align}
&\lim_{\SNR\rightarrow\infty}\mi{|X_1|^2}{V}
\ge \lim_{\SNR\rightarrow\infty} \bigg\{ -\frac{3}{2}\frac{1 }{\SNR^{1-\alpha(t+1)}-1}\nonumber\\
&+\frac{1}{2}\ln\left((\SNR-\SNR^{\alpha (t+1)})^2+\rho(\SNR^{1-\alpha(1-\beta)}-\SNR^{\alpha (t+\beta)})\right) \nonumber\\
&- \frac{1}{2}\ln\left(\pi\rho\SNR^{1+\alpha(\beta-1)}\left(1-\frac{1 }{\SNR^{1-\alpha(t+1)}}\right)\right)\nonumber\\
& - \rho^{-1}\SNR^{\alpha(1-\beta)}\left(\SNR^{1-3\alpha}\cdot \frac{\gamma^2}{45}+  2+\SNR^{-\alpha t}\right)\bigg\}\nonumber\\
&= \lim_{\SNR\rightarrow\infty} \bigg\{-\frac{3}{2}\frac{1 }{\SNR^{1-\alpha(t+1)}-1}\nonumber\\
&+\frac{1}{2}\ln\left(\frac{(\SNR-\SNR^{\alpha (t+1)})^2}{\SNR^{1+\alpha(\beta-1)}}+\rho\frac{\SNR^{1-\alpha(1-\beta)}-\SNR^{\alpha (t+\beta)}}{\SNR^{1+\alpha(\beta-1)}}\right) \nonumber\\
&\quad- \frac{1}{2}\ln\left(\pi\rho\left(1-\frac{1 }{\SNR^{1-\alpha(t+1)}}\right)\right)\nonumber\\
& - \rho^{-1}\SNR^{\alpha(1-\beta)}\left(\SNR^{1-3\alpha}\cdot \frac{\gamma^2}{45}+  2+\SNR^{-\alpha t}\right)\bigg\}\nonumber\\
&= \lim_{\SNR\rightarrow\infty} \bigg\{ -\frac{3}{2}\frac{1 }{\SNR^{1-\alpha(t+1)}-1}\nonumber\\
&+\frac{1}{2}\ln\left(\SNR^{1+\alpha(1-\beta)}-2\SNR^{\alpha (t-\beta+2)}+\SNR^{\alpha (2t-\beta+3)-1}\right.\nonumber\\
&\qquad\left.+\rho-\frac{\rho}{\SNR^{1-\alpha(t+1)}}\right) \nonumber\\
&\quad- \frac{1}{2}\ln\left(\pi\rho\left(1-\frac{1 }{\SNR^{1-\alpha(t+1)}}\right)\right) \nonumber\\
&\quad- \rho^{-1}\SNR^{\alpha(1-\beta)}\left(\SNR^{1-3\alpha}\cdot \frac{\gamma^2}{45}+  2+\SNR^{-\alpha t}\right)\bigg\}.
\end{align}
In order to have a tight bound, we need to satisfy the constraints
\begin{equation}
 \left\{\begin{array}{l}
         1-\alpha(t+1) > 0 \\
         \alpha(1-\beta) + 1-3\alpha \le 0 \\
         \alpha(1-\beta) \le 0 \\
         \alpha(1-\beta) -\alpha t \le 0 \\
         1+\alpha(1-\beta) > \alpha (t-\beta+2)
        \end{array}\right.
\end{equation}
that reduce to
\begin{equation}
 \left\{\begin{array}{l}
         \alpha < 1/(t+1) \\
         \alpha(1-\beta) + 1-3\alpha \le 0 \\
         \alpha(1-\beta) \le 0.
        \end{array}\right.
\end{equation}
Next we consider the two cases $1/3<\alpha<1$ and $0<\alpha<1/3$.

If $\alpha> 1/3$, i.e., $1-3\alpha<0$, then we have to satisfy
\begin{equation}
 \left\{\begin{array}{l}
         \alpha < 1/(t+1) \\
         \beta \ge 1.
        \end{array}\right.
\end{equation}
to get
\begin{align}
\lim_{\SNR\rightarrow\infty}&\left\{\mi{|X_1|^2}{V}-\frac{1+\alpha(1-\beta)}{2}\ln\left(\SNR\right)\right\}\nonumber\\
&\quad\ge  - \frac{1}{2}\ln\left(\pi\rho\right) -  \frac{2}{\rho},
\end{align}
and the tightest bound is obtained by choosing $\beta=1$ and $\rho=4$:
\begin{align}
\lim_{\SNR\rightarrow\infty}\left\{\mi{|X_1|^2}{V}-\frac{1}{2}\ln\left(\SNR\right)\right\}
&\ge  - \frac{1}{2}\ln\left(4\pi e\right).
\end{align}

If $\alpha< 1/3$, i.e., $1-3\alpha>0$, then we have to satisfy
\begin{equation}
 \left\{\begin{array}{l}
         \alpha < 1/(t+1) \\
         \alpha  \ge 1/(\beta+2).
        \end{array}\right.
\end{equation}
By choosing $\beta = \alpha^{-1}-2$ and $\rho = 2\gamma^2/45$ we get
\begin{align}
\lim_{\SNR\rightarrow\infty}\left\{\mi{|X_1|^2}{V}-\frac{3\alpha}{2}\ln\left(\SNR\right)\right\}
&\ge  - \frac{1}{2}\ln\left(\frac{2\pi\gamma^2 e}{45}\right).
\end{align}

\section{A lower bound to $\expect{\cos(\angle(\rho+W))}$}\label{App:C}
The pdf of $\Psi=\angle(\rho + W)$ is~\cite{Aldis}
\begin{equation}\label{app:pdf}
 p_{\Psi}(\psi) = \frac{1}{2\pi}e^{-\rho^2}+\frac{1}{\sqrt{4\pi}}\rho\cos(\psi) e^{-\rho^2\sin^2(\psi)} \erfc(-\rho\cos(\psi))
\end{equation}
where $\erfc:x\mapsto 1-\erf(x)$ is the complementary error function. A lower bound to $\expect{\cos(\Psi)}$ is\footnote{The proof is the one proposed in the Ph.D. thesis of H. Ghozlan.}:
\begin{align}
 &\expect{\cos(\Psi)} \stackrel{(a)}{=} 2\int_0^\pi \cos(\psi) p_{\Psi}(\psi) \diff \psi \nonumber\\
 &\stackrel{(b)}{=} \int_0^\pi \frac{\rho}{\sqrt{\pi}}\cos^2(\psi) e^{-\rho^2\sin^2(\psi)} \erfc(-\rho\cos(\psi)) \diff \psi \nonumber\\
 &\stackrel{(c)}{\ge} \int_0^{\pi/2} \frac{\rho}{\sqrt{\pi}}\cos^2(\psi) e^{-\rho^2\sin^2(\psi)} \erfc(-\rho\cos(\psi)) \diff \psi \nonumber\\
 &\stackrel{(d)}{\ge} \int_0^{\pi/2} \frac{\rho}{\sqrt{\pi}}\cos^2(\psi) e^{-\rho^2\sin^2(\psi)} (2-e^{-\rho^2\cos^2(\psi)}) \diff \psi \nonumber\\
 &= \int_0^{\pi/2} \frac{2\rho}{\sqrt{\pi}}\cos^2(\psi) e^{-\rho^2\sin^2(\psi)} \diff \psi - \frac{\sqrt{\pi}}{4} \rho e^{-\rho^2} \label{app:integral}
\end{align}
where~$(a)$ follows by symmetry, $(b)$ follows by using~\eqref{app:pdf} and $\int_0^\pi \cos(\psi) \diff \psi = 0$, $(c)$ holds because the integrand is non-negative over the interval $[\pi^2,\pi]$,
 $(d)$ holds because $\erfc(\rho\cos(\psi))\ge 2-e^{-\rho^2\cos^2(\psi)}$ and $\cos^2(\psi)e^{-\rho^2\sin^2(\psi)}\ge 0$, and finally the last step follows by direct integration.

 We bound the integral in~\eqref{app:integral} as follows
\begin{align}
 \int_0^{\pi/2} \frac{2\rho}{\sqrt{\pi}}&\cos^2(\psi) e^{-\rho^2\sin^2(\psi)} \diff \psi \nonumber\\
 &\stackrel{(a)}{\ge} \frac{2\rho}{\sqrt{\pi}}\int_0^{\pi/2} (1-\psi^2)e^{-\rho^2\psi^2} \diff \psi \nonumber\\
 &= \left(1-\frac{1}{2\rho^2}\right)\erf\left(\frac{\pi\rho}{2}\right)  +\frac{1}{2\sqrt{\pi}\rho} e^{-\pi^2 \rho^2/4} \nonumber\\
 &\ge \erf\left(\frac{\pi\rho}{2}\right)  -\frac{1}{2\rho^2}
\end{align}
where inequality~$(a)$ follows from $\cos^2(\psi)e^{-\rho^2\sin^2(\psi)} \ge (1-\psi^2)e^{-\rho^2\psi^2}$, and the last step holds because $\erf(\cdot)\le 1$ and the last term is non-negative.
Substituting back into~\eqref{app:integral} yields
\begin{align}
 \expect{\cos(\Psi)} &\ge \erf\left(\frac{\pi\rho}{2}\right)  -\frac{1}{2\rho^2} - \frac{\sqrt{\pi}}{4} \rho e^{-\rho^2} \nonumber\\
 &\stackrel{(a)}{\ge} 1-e^{-\pi^2\rho^2/4}  -\frac{1}{2\rho^2} - \frac{\sqrt{\pi}}{4} \rho e^{-\rho^2} \nonumber\\
 &\stackrel{(b)}{\ge} 1- \frac{4}{\pi^2 \rho^2 e}  -\frac{1}{2\rho^2} - \frac{\sqrt{\pi}}{4\rho^2} \left(\frac{3}{2e}\right)^{3/2} \nonumber\\
 &\ge 1- \frac{1}{\rho^2} 
\end{align}
where step~$(a)$ is due to $\erfc(x)\le e^{-x^2}$, and inequality~$(b)$ follows from $\rho^3 e^{-\rho^2}\le (3/(2e))^{3/2}$ and $e^{-\pi^2 \rho^2/4} \le 4/(\pi^2 \rho^2 e)$.

\section{An upper bound to $\expect{|F_1|^{-2}}$}\label{App:D}

Denoting $Z=|F_1|$, we compute an upper bound as follows
\begin{align}
  &\expect{Z^{-2}} = \lim_{\delta\downarrow 0}\int_\delta^\infty \frac{1}{x^2} p_{Z}(x) \diff x \nonumber\\
  &\stackrel{(a)}{=} \lim_{\delta\downarrow 0}\left\{-\frac{\Pr(Z\le \delta)}{\delta^2} + \int_\delta^\infty \frac{2}{x^3} \Pr(Z\le x) \diff x \right\}\nonumber\\
  &\stackrel{(b)}{\le} \int_0^\infty \frac{2}{x^3} \expect{1(Z\le x)} \diff x \nonumber\\
  &= \int_0^\varepsilon \frac{2}{x^3} \expect{1(Z\le x)} \diff x +\int_\varepsilon^\infty \frac{2}{x^3} \expect{1(Z\le x)} \diff x\nonumber\\
  &\le \int_0^\varepsilon \frac{2}{x^3} \expect{g(Z)} \diff x +\frac{1}{\varepsilon^2} \label{app:negmoment}
\end{align}
where $\varepsilon$ is a suitably chosen positive number, step~$(a)$ follows by integrating by parts, inequality~$(b)$ holds because the cumulative function is always positive, 
and the last inequality holds by choosing a function
$g(Z)\ge 1(Z\le x)$ for the first integral (i.e., for $0< x < \varepsilon$) and the inequality $\expect{1(Z\le x)}\le 1$ for the second integral, that can be computed in closed form.

As for the function $g(Z)$ we choose
\begin{align}
 &g(Z) = a(1-Z^2)\left(1-Z^2\rho_1\right)\left(1-Z^2\rho_2\right) \nonumber\\
 &= a\left[1-Z^2\left(1+\rho_1+\rho_2\right) +Z^4 (\rho_1+\rho_2+\rho_1\rho_2) -Z^6 \rho_1\rho_2 \right], \label{app:f}
\end{align}
a polynomial whose positive roots are $\{1,\rho_1^{-1/2},\rho_2^{-1/2}\}$, and with $g(0)=a>1$. To guarantee the positivity of $g(Z)$ for $0\le Z\le 1$,
we can design the roots $\rho_1^{-1/2}$ and $\rho_2^{-1/2}$ to be greater than 1, hence $\rho_1$ and $\rho_2$ less than 1.

If the polynomial $g(Z)$ satisfies the condition $\expect{g(Z)}=0$, then we have a finite bound in~\eqref{app:negmoment}:
\begin{equation}
 \expect{Z^{-2}} \le \frac{1}{\varepsilon^2}.\label{app:bound}
\end{equation}
Imposing the condition $\expect{g(Z)}=0$ in~\eqref{app:f} gives
\begin{equation}\label{app:rho2}
 \rho_2 = \frac{-1+\expect{Z^2}(1+\rho_1)-\expect{Z^4}\rho_1}{-\expect{Z^2}+\expect{Z^4}(1+\rho_1)-\expect{Z^6}\rho_1}.
\end{equation}
We want $\rho_2$ to be positive, for this we distinguish two cases. In the first case we have both numerator and denominator of~\eqref{app:rho2} positive, and this is satisfied if
\begin{equation}
 \rho_1 \ge \max\left\{\frac{\expect{1-Z^2}}{\expect{Z^2(1-Z^2)}},\frac{\expect{Z^2(1-Z^2)}}{\expect{Z^4(1-Z^2)}}\right\}\ge 1,
\end{equation}
so this situation is not wanted. The other case is where both numerator and denominator are negative, i.e., for
\begin{equation}\label{app:rho_1_1}
 \rho_1 \le \min\left\{\frac{\expect{1-Z^2}}{\expect{Z^2(1-Z^2)}},\frac{\expect{Z^2(1-Z^2)}}{\expect{Z^4(1-Z^2)}}\right\}.
\end{equation}
Moreover, we want $\rho_2 \le 1$, and imposing this condition on~\eqref{app:rho2} when numerator and denominator are negative means
\begin{equation}\label{app:rho_1_2}
 \rho_1 \ge \frac{\expect{(1-Z^2)^2}}{\expect{Z^2(1-Z^2)^2}}
\end{equation}
Conditions~\eqref{app:rho_1_1} and~\eqref{app:rho_1_2} can be numerically checked by considering that~\cite{GhozlanISIT2013}
\begin{equation}
 \expect{Z^2} = \frac{2}{\alpha^2} \left(- 1 + e^{-\alpha t} +\alpha t  \right)
\end{equation}
\begin{align}
 \expect{Z^4} = \frac{1}{\alpha^4} &\left(\frac{87}{2}-\frac{392}{9}e^{-\alpha } + \frac{1}{18} e^{-4\alpha } - 30\alpha  + 8\alpha^2  \right. \nonumber\\
 & \quad\left. - \frac{40}{3} \alpha  e^{-\alpha } \right)
\end{align}
\begin{align}
 \expect{Z^6} = \frac{1}{\alpha^6} &\left(-100 \alpha^3  e^{-\alpha } + 144 \alpha^3   - \frac{3}{25}\alpha  e^{-4\alpha }     \right. \nonumber\\
&\quad - \frac{11991}{8} \alpha  e^{-\alpha }  + 1499 \alpha - \frac{1}{200} \alpha  e^{-9\alpha } \nonumber\\
&\quad \left. -\frac{2123}{3} \alpha^2  e^{-\alpha } + \frac{4}{15} \alpha^2 e^{-4\alpha } - 792 \alpha^2  \right)
\end{align}
where $\alpha = \gamma^2\Delta/2$. For example, for $a=1.3$ and $\gamma \sqrt{\Delta}= 0.1$ the roots are $\rho_1^{-1/2}\gg 1, \rho_2^{-1/2}\approx 1.0001>1$.

The last thing to check is that 
\begin{equation}\label{app:lastcondition}
 g(Z)\ge 1, \text{ for } 0\le Z\le x \text{ and all } 0\le x \le \varepsilon.
\end{equation}
Studying the convexity of $g(Z)$, we find that the function is concave in $0\le Z\le  \varepsilon_{\cap}$ with
\begin{equation}
 \varepsilon_{\cap}^2= \frac{\rho_1+ \rho_2+\rho_1 \rho_2}{5\rho_1 \rho_2}-\sqrt{\left(\frac{\rho_1+ \rho_2+\rho_1 \rho_2}{5\rho_1 \rho_2}\right)^2-\frac{1+\rho_1+\rho_2}{15\rho_1 \rho_2}}.
\end{equation}
Moreover, we have $g(\varepsilon_1)=1$ with $\varepsilon_1^2$ given by Cardano's formula.
 Condition~\eqref{app:lastcondition} is verified if $\varepsilon\le \min\{ \varepsilon_1, \varepsilon_{\cap}\}$, because for all $0\le Z \le \epsilon$ we can guarantee 
 $g(Z)\ge 1$ thanks to $g(0)\ge 1$, $g(\epsilon)\ge 1$, and concavity for $g(Z)$. For example, for $a=1.3$ and $\gamma \sqrt{\Delta}= 0.1$ we have $\varepsilon_1 \approx 0.3506$ and 
 $\varepsilon_{\cap}\approx 0.5774$, so we choose $\varepsilon = 0.3506$. Using~\eqref{app:bound} this gives a bound $\expect{|F|^{-2}}\le 8.1353$ for all $\gamma \sqrt{\Delta}\le 0.1$.

\end{document}